\documentclass[pre,letterpaper,twocolumn,floatfix]{revtex4}
\usepackage{graphicx,amssymb}

\begin{document}

\title{On the extensivity of the entropy $S_q$ for $N\le 3$ 
specially correlated binary subsystems}

\author{Yuzuru Sato$^a$ and
Constantino Tsallis$^{a,b}$ \footnote{ysato@santafe.edu, tsallis@santafe.edu}
}

\address{$^a$Santa Fe Institute,
1399 Hyde Park Road,
Santa Fe, New Mexico 87501,  USA \\
and\\
$^b$Centro Brasileiro de Pesquisas Fisicas, Rua Xavier Sigaud 150, 
22290-180 Rio de Janeiro-RJ, Brazil
}
\date{\today}
\begin{abstract}
Many natural and artificial systems whose range of interaction is 
long enough are known  to exhibit (quasi)stationary  states that 
defy the standard, Boltzmann-Gibbs statistical mechanical prescriptions. For handling such 
anomalous systems (or at least some classes of them), {\it nonextensive} statistical mechanics 
has been proposed based on the entropy $S_{q}\equiv
 k\,(1-\sum_{i=1}^Wp_i^{\;q})/(q-1)$, with  
$S_1=-k\Sigma_{i=1}^{W} p_i \ln p_i$ (Boltzmann-Gibbs entropy). 
Special collective correlations can be mathematically constructed 
such that the strictly {\it additive} entropy is now $S_q$ 
for an adequate value of $q \ne 1$, whereas Boltzmann-Gibbs entropy 
is {\it nonadditive}. Since important classes of systems exist 
for which the strict additivity of Boltzmann-Gibbs entropy is replaced 
 by asymptotic additivity (i.e., extensivity), a variety of 
 classes are expected to exist for which the strict additivity of $S_q\;\;(q\ne
 1)$ is similarly replaced by asymptotic additivity (i.e.,
 extensivity). All probabilistically well defined systems whose adequate entropy is
 $S_{1}$ are called {\it extensive} (or {\it normal}). They correspond to a number $W^{\mbox{\it eff}}$ of {\it effectively} occupied states which grows {\it exponentially} with the number $N$ of elements (or subsystems). Those whose adequate entropy is $S_q \;\;(q \ne 1)$ 
are currently called {\it nonextensive} (or {\it anomalous}). They correspond to  $W^{\mbox{\it eff}}$ growing like a {\it power} of $N$. To illustrate this scenario, recently addressed \cite{tsallisSF},  
we provide in this paper details about systems composed by $N=2,3$ 
two-state subsystems. 
\end{abstract}
\maketitle

\section{Introduction}
\label{section_introduction}

It is well known that, if we have a system composed by $N$
 statistically independent subsystems (i.e., such that all joint
 probabilities factorize into the marginal ones corresponding to each
 subsystem), the Boltzmann-Gibbs (BG) entropy $S_{BG} \equiv\ -k
 \sum_{i=1}^W p_i \ln p_i $ is strictly {\it additive}, i.e.,
 $S_{BG}(A_1+A_2+...+A_N)= S_{BG}(A_1)+S_{BG}(A_2)+...+S_{BG}(A_N)$. A
 plethora of physical systems is known for which this remarkable
 property still holds {\it asymptotically} ($N \to\infty$). Such is
 the case, for instance, of virtually all many-body Hamiltonian
 systems involving {\it short}-range two-body interactions. This
 property is called {\it extensivity}, adopting the language of
 thermodynamics, where it plays an important role. Many natural and
 artificial systems exist however that do {\it not} belong to this
 class, such as many-body Hamiltonian systems involving two-body
 interactions  whose range of interaction is {\it long} enough
 (Newtonian gravitation is one famous example). Such systems are known
 to exhibit stationary (or {\it quasi-stationary} or {\it metastable})
 states that defy the usual, BG statistical mechanical prescriptions. For
 handling at least some of such anomalous systems, a generalization of
 BG statistical mechanics, has been proposed in 1988 \cite{Tsallis88}, which is usually
 referred to as {\it nonextensive} statistical mechanics (see \cite{SalinasTsallis} for reviews). It is based
 on the entropy $S_{q}\equiv
 k\,(1-\sum_{i=1}^Wp_i^{\;q})/(q-1)\;\;(q\in{\mathbb
 R}\,;\;S_1=S_{BG})$, which generalizes the BG one. It has been shown
 recently that special collective correlations can be mathematically
 constructed such that the entropy which is strictly {\it additive} is
 now $S_q$ for an adequate value of $q \ne 1$ (directly determined by
 the type of correlations), whereas $S_{BG}$ is {\it nonadditive}. It
 is easy to imagine that, in the same way that important classes of
 systems exist for which the strict additivity of $S_{BG}$ is replaced
 by just asymptotic additivity (i.e., extensivity), a variety of
 classes must exist for which the strict additivity of $S_q\;\;(q\ne
 1)$ is similarly replaced by asymptotic additivity (i.e.,
 extensivity). Such systems would be the object of the so-called
 nonextensive statistical mechanics. Then, as a kind of bizarre
 linguistic twist, it turns out that the appropriate entropy $S_q$ for
 such, so-called {\it nonextensive} systems, is in fact expected to be
 {\it extensive}. The generic scenario is therefore as follows: all probabilistically well defined systems are expected to have an entropy which is extensive; those whose appropriate entropy is $S_{BG}$ (or its associated forms, such as those adapted to fermions and bosons) are called {\it extensive}, and those whose appropriate entropy is $S_q \;\;(q \ne 1)$ (or even some other entropic form) are called {\it nonextensive}. 

A quantity $X(A)$ associated with a system $A$ is said {\it additive} (see \cite{tsallisSF}, which we closely follow here) with regard to a specific composition of $A$ and $B$ if it satisfies 
\begin{equation}
X(A+B)=X(A)+X(B) \;,
\end{equation}
where $+$ inside the argument of $X$ precisely indicates that composition. 

If, instead of two subsystems $A$ and $B$, we have $N$ of them ($A_1, A_2, ..., A_N$), then we have that
\begin{equation}
X(\sum_{i=1}^N A_i)=\sum_{i=1}^N X(A_i) \;.
\end{equation}
If the subsystems (e.g., just the elements of the full system) happen to be all equal (a quite common case), then we have that
\begin{equation}
X(N)=NX(1)\;,
\end{equation}
with the notations $X(N) \equiv X(\sum_{i=1}^N A_i)$ and $X(1) \equiv X(A_1)$. 

An intimately related concept is that of {\it extensivity}. It appears frequently in thermodynamics and elsewhere, and corresponds to a weaker demand, namely that of
\begin{equation}
\lim_{N \to\infty}\frac{|X(N)|}{N} < \infty \,.
\end{equation}
Clearly, all quantities that are additive with regard to a given composition law, also are extensive with regard to that same composition (and $\lim_{N \to\infty}X(N)/N=X(1)$), whereas the opposite is not necessarily true. Let us apply these remarks to entropy.

Boltzmann-Gibbs ($BG$) statistical mechanics is based on the entropy
\begin{equation}
S_{BG} \equiv\ -k \sum_{i=1}^W p_i \ln p_i \;,
\end{equation}
with
\begin{equation}
\sum_{i=1}^W p_i=1 \;,
\end{equation}
where $p_i$ is the probability associated with the $i^{th}$ microscopic state of the system, and $k$ is Boltzmann constant. In the particular case of equiprobability, i.e., $p_i=1/W$  $(\forall i)$, Eq. (5) yields the well known {\it Boltzmann principle}:
\begin{equation}
S_{BG}=k \ln W \;.
\end{equation}
From now on, and without loss of generality, we shall take $k$ equal to unity.  

Nonextensive statistical mechanics
 is based on the so-called ``nonextensive" entropy $S_q$ defined as follows:
\begin{equation}
S_{q}\equiv\frac{1-\sum_{i=1}^Wp_i^{\;q}}{q-1}\;\;\;
(q\in{\mathbb R};\;S_1=S_{BG}) \;.
\label{q_entropy}
\end{equation}
(Later on we come back onto the denomination ``nonextensive").

For equiprobability (i.e., $p_i=1/W,\,\forall i$), Eq. (8) yields
\begin{equation}
S_q=\ln_q W \;,
\end{equation}
with the {\it $q$-logarithm} function defined  as
\begin{equation}
\ln_q z \equiv \frac{z^{1-q}-1}{1-q} \;\;\;(z \in{\mathbb R}; \;z>0; \;\ln_1 z=\ln z) \;.
\end{equation}
The inverse function, the {\it $q$-exponential}, is given by
\begin{equation}
e_q^z \equiv [1+(1-q)z]^{1/(1-q)} \;\;\;(e_1^z=e^z) 
\end{equation}
if the argument $1+(1-q)z$ is positive, and vanishes otherwise.
Following a common usage, we shall from now on cease distinguishing between {\it additive} and {\it extensive}, and use exclusively the word {\it extensive} in the sense of either strictly or only asymptotically additive. 

\section{$N$ subsystems}

\subsection{General considerations}

Consider a system composed by $N$ subsystems $A_1,A_2,...,A_N$ having respectively $W_{A_1},W_{A_2},...,W_{A_N}$ possible microstates (we only address here the basic case of {\it discrete} microstates). The total number of possible microstates for the system $A_1+A_2+...+A_N$ is then {\it in principle} $W \equiv W_{A_1+A_2+...+A_N}=W_{A_1}W_{A_2}...W_{A_N}$. We emphasized the expression ``in principle" because we shall see that a more or less severe reduction of the full phase space might occur in the presence of a special type of strong correlations between the subsystems. 

We shall use the notation $p_{ij}^{A_1+A_2+...+A_N} \;\;(i_1=1,2,...,W_{A_1};\; i_2=1,2,...,W_{A_2};...)$ for the {\it joint probabilities}, hence
\begin{equation}
\sum_{i_1=1}^{W_{A_1}}\sum_{i_2=1}^{W_{A_2}}... \sum_{i_N=1}^{W_{A_N}}p_{i_1i_2...i_N}^{A_1+A_2+...+A_N} =1 \;.
\end{equation}
The $A_1-${\it marginal probabilities} are defined as follows:
\begin{equation}
p_{i_1}^{A_1} \equiv \sum_{i_2=1}^{W_{A_2}}... \sum_{i_N=1}^{W_{A_N}}p_{i_1i_2...i_N}^{A_1+A_2+...+A_N} \,,
\end{equation}
hence
\begin{equation}
\sum_{i_1=1}^{W_{A_1}} p_{i_1}^{A_1} =1 \,.
\end{equation}
Analogously are defined all the other $N-1$ {\it one-subsystem}$-${\it marginal probabilities}.
The $A_1A_2-${\it marginal probabilities} are defined as follows:
\begin{equation}
p_{i_1i_2}^{A_1+A_2} \equiv \sum_{i_3=1}^{W_{A_3}}... \sum_{i_N=1}^{W_{A_N}}p_{i_1i_2...i_N}^{A_1+A_2+...+A_N} \,,
\end{equation}
hence
\begin{equation}
\sum_{i_1=1}^{W_{A_1}} \sum_{i_2=1}^{W_{A_2}}p_{i_1i_2}^{A_1+A_2} =1 \,.
\end{equation}
Similarly are defined all the other $[N(N-1)/2]-1$ {\it
two-subsystem}$-${\it marginal probabilities}, as well as all the
other {\it many-subsystem}$-${\it marginal probabilities}. The most
general $N=2$ case is indicated in Table I.

\begin{table}[htbp]
\begin{center}
\begin{tabular}{c||c|c|c|c||c}
 $_A\setminus^B$    &  1                          & 2                            &$\;\;\;\;...\;\;\;\;$         &$W_B$                                   \\[1mm] \hline\hline
1           &  $\;\;p_{11}^{A+B}\;\;$     & $\;\;p_{12}^{A+B}\;\;$       &...                     & $\;\;p_{1W_B}^{A+B}\;\;$                                             & $\;\;p_1^A\;\;$   \\[3mm] \hline
2           &  $p_{21}^{A+B}$             & $p_{22}^{A+B}$               &...        & $\;\;p_{2W_B}^{A+B}\;\;$                                              & $p_2^A$   \\[3mm] \hline
...          &  $\;\;...\;\;$                        &  $\;\;...\;\;$                        &  $\;\;...\;\;$       &  $\;\;...\;\;$                                                                   &  $\;\;...\;\;$    \\[3mm] \hline   
$W_A$  &$\;\;p_{W_A1}^{A+B}\;\;$  &$\;\;p_{W_A2}^{A+B}\;\;$  &  $\;\;...\;\;$       &$\;\;p_{W_AW_B}^{A+B}\;\;$        &$p_{W_A}^A$                    \\[3mm] \hline \hline
             &  $p_1^B$                         & $p_2^B$                          &...                     &$p_{W_B}^B$                                                                   & 1
\end{tabular}
\end{center}
 \label{tab:jp2s}
\caption{Joint probabilities for two subsystems}
\end{table}

The central point that the present paper addresses is whether it is or not possible to satisfy all this specific structure of joint and marginal probabilities, and {\it simultaneously} impose the condition
\begin{equation}
S_q(A_1+A_2+...+A_N)=S_q(A_1)+S_q(A_2)+...+S_q(A_N) \,,
\end{equation}
where $S_q(A_1+A_2+...+A_N)$ is calculated with the joint probabilities, and $S_q(A_r)$ is calculated with the $A_r-${\it marginal probabilities} ($r=1,2,...,N$). To simultaneously satisfy, in fact, not only Eq. (17) but also
\begin{eqnarray}
S_q(A_1+A_2+...+A_N) \nonumber \\
 = [S_q(A_1+A_2)+S_q(A_1+A_3)     \nonumber \\
+...+S_q(A_{N-1}+A_N)]/(N-1) \,,
\end{eqnarray}
and all similar ones calculated with all possible combinations of {\it many-subsystem}$-${\it marginal probabilities}. It is kind of trivial that at least one solution exists, namely that of {\it mutually independent} subsystems with the entropy $S_{BG}$. Indeed, we can verify that the hypothesis
\begin{equation}
p_{i_1i_2...i_N}^{A_1+A_2+...+A_N}=\prod_{r=1}^N p_{i_r}^{A_r} \;\;\;(\forall (i_1,i_2,...,i_N)) 
\end{equation}
implies Eqs. (17), (18) and all the similar ones. 
But it is {\it by no means} trivial that a different choice is possible involving collective correlations and a value of $q$ different from unity. Paper \cite{tsallisSF} answered affirmatively {\it precisely} to this question.  We shall present here details of such solutions, namely for $N=2$ and $N=3$ binary subsystems. 

\subsection{$N=2$ specially correlated binary systems} 

The most general  joint probabilities for two binary subsystems (noted $A$ and $B$, with $W_A=W_B=2$) are indicated in Table II.

\begin{table}[htbp]
\begin{center}
\begin{tabular}{c||c|c||c}
$_A\setminus^B$    &  1                          & 2                                                    \\[1mm] \hline\hline
1  &  $\;\;p_{11}^{A+B}\;\;$   & $\;\;p_{12}^{A+B}\;\;$   & $\;\;p_1^A\;\;$   \\[3mm] \hline
2  &  $p_{21}^{A+B}$           & $p_{22}^{A+B}$           & $p_2^A$   \\[3mm] \hline \hline
   &  $p_1^B$                       & $p_2^B$                       & 1

\end{tabular}
\end{center}
 \label{tab:jp2bs}
\caption{Joint probabilities for two binary subsystems}
\end{table}

It trivially verifies Eq. (17) with $q=1$ in Table III ({\it top})
and it was shown in \cite{tsallisSF} that the Table III 
({\it middle}) satisfies Eq. (17) with $q=0$ with $p_1^A+p_1^B>1$.  
It is possible to interpolate between Table III ({\it top}) and Table
III ({\it middle}) through Table III ({\it bottom}) ($0 \le q \le 1$),  
where the function $f_q(x,y)$ is defined as follows: 
\begin{eqnarray}
(p_1^A)^q+(1-p_1^A)^q+(p_1^B)^q+(1-p_1^B)^q    \nonumber \\
-[f_q(p_1^A,p_1^B)]^q     \nonumber \\
-[p_1^A-f_q(p_1^A,p_1^B)]^q  
-[p_1^B-f_q(p_1^A,p_1^B)]^q   \nonumber \\
-[1-p_1^A-p_1^B + f_q(p_1^A,p_1^B)]^q=1
\end{eqnarray}
We verify that
\begin{eqnarray}
f_q(p_1^A,p_1^B)&=& f_q(p_1^B,p_1^A) ,\nonumber\\
f_q(p,1)&=&p ,\nonumber\\
f_1(p_1^A,p_1^B)&=&p_1^Ap_1^B ,\nonumber\\
 f_0(p_1^A,p_1^B)&=&p_1^A+p_1^B-1 , 
\end{eqnarray}
and also Eq. (17). Typical examples of the function $f_q(x,y)$ are shown in Fig. 1.

\begin{table}[htbp]
\begin{center}
\begin{tabular}{c||c|c||c}
$_A\setminus^B$    &  1                          & 2                                                    \\[1mm] \hline\hline
1  &  $\;\;p_1^Ap_1^B\;\;$   & $\;\;p_1^Ap_2^B\;\;$   & $\;\;p_1^A\;\;$   \\[3mm] \hline
2  &  $p_2^Ap_1^B$           & $p_2^Ap_2^B$           & $p_2^A$   \\[3mm] \hline \hline
   &  $p_1^B$                       & $p_2^B$                       & 1

\end{tabular}
\end{center}

\begin{center}
\begin{tabular}{c||c|c||c}
 $_A\setminus^B$    &  1                          & 2                                                    \\[1mm] \hline\hline
1  &  $\;\;p_1^A+p_1^B-1\;\;$   & $\;\;1-p_1^B\;\;$   & $\;\;p_1^A\;\;$   \\[3mm] \hline
2  &  $1-p_1^A$                       & 0                           & $1-p_1^A$   \\[3mm] \hline \hline
   &  $p_1^B$                           & $1-p_1^B$                    & 1
\end{tabular}
\end{center}

\begin{center}
\begin{tabular}{c||c|c||c}
 $_A\setminus^B$    &  1                          & 2                                                    \\[1mm] \hline\hline
1  &  $\;\;f_q(p_1^A,p_1^B)\;\;$              & $\;\;p_1^A-f_q(p_1^A,p_1^B)\;\;$            & $\;\;p_1^A\;\;$   \\[3mm] \hline
2  &  $p_1^B-f_q(p_1^A,p_1^B)$                   & $1-p_1^A-p_1^B+f_q(p_1^A,p_1^B)$              & $1-p_1^A$   \\[3mm] \hline \hline
   &  $p_1^B$                               & $1-p_1^B$                           & 1
\end{tabular}
\end{center}
 \label{tab:jpsc2bs}
\caption{Joint probabilities for independent two binary subsystems
 ({\it top}), for specially correlated two binary subsystems ({\it middle}), 
and for an interpolation between them ({\it bottom}).}
\end{table}

\begin{figure}[htbp]
\begin{center}
\hspace{3mm}
\includegraphics[scale=0.7]{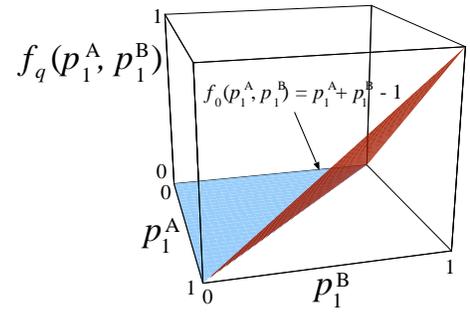}
\end{center}

\begin{center}
\includegraphics[scale=0.7]{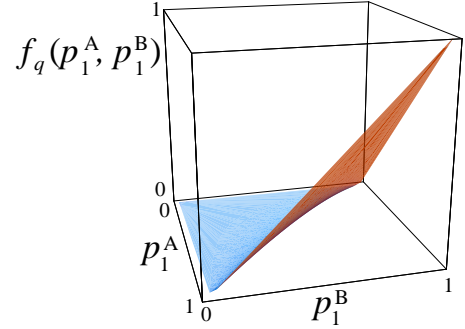}
\end{center}

\begin{center}
\includegraphics[scale=0.7]{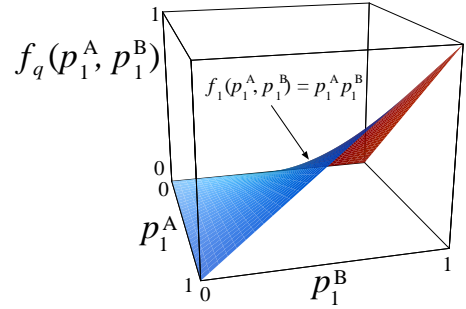}
\end{center}
\label{fig:2surface}
\caption{Typical surfaces for $N=2$ case: ~$q=0$ ({\it top}), $q=0.5$
 ({\it middle}; the graphically unperfect match at  $p_1^B=0$ and at $p_1^A=p_1^B=1$ is only due to numerical imprecision), and $q=1$ ({\it bottom}).}
\end{figure} 

Let us consider the simple case where $A=B$ (hence $p_1^A=p_1^B \equiv
p$). Tables III become respectively Tables IV, where $f_q(p) \equiv f_q(p,p)$  
satisfies the relation 
\begin{equation} 
2p^q+2(1-p)^q-(f_q)^q-2(p-f_q)^q-(1-2p+f_q)^q=1\,.
\end{equation}   
In Fig. 2 we present $f_q(p)$ for typical
values of $q$ and the $q-$dependence of $f_q(1/2)$. It can be straighforwardly verified that $S_q(2)=2S_q(1)$ (see Fig. 3).

\begin{table}[htbp]
\begin{center}
\begin{tabular}{c||c|c||c}
 $_A\setminus^B$    &  1                          & 2                                                    \\[1mm] \hline\hline
1  &  $\;\;p^2\;\;$                                       & $\;\;p(1-p)\;\;$                         & $\;\;p\;\;$  
 \\[3mm] \hline
2  &  $p(1-p)$                                           & $(1-p)^2$                                & $1-p$   
\\[3mm] \hline \hline
    &  $p$                                                  & $1-p$                                       & 1
\end{tabular}
\end{center}

\begin{center}
\begin{tabular}{c||c|c||c}
 $_A\setminus^B$    &  1                          & 2                                                    \\[1mm] \hline\hline
1  &  $\;\;2p-1\;\;$                 & $\;\;1-p\;\;$   & $\;\;p\;\;$   \\[3mm] \hline
2  &  $1-p$                           & 0                  & $1-p$   \\[3mm] \hline \hline
    &  $p$                              & $1-p$           & 1
\end{tabular}
\end{center}

\begin{center}
\begin{tabular}{c||c|c||c}
 $_A\setminus^B$    &  1                          & 2                                                    \\[1mm] \hline\hline
1  &  $\;\;f_q(p)\;\;$              & $\;\;p-f_q(p)\;\;$            & $\;\;p\;\;$   \\[3mm] \hline
2  &  $p-f_q(p)$                   & $1-2p+f_q(p)$              & $1-p$   \\[3mm] \hline \hline
   &  $p$                               & $1-p$                           & 1
\end{tabular}
\end{center}
\label{tab:jpsc2bss}
\caption{Joint probabilities for independent two binary subsystems
 ({\it top}) , for specially correlated two binary subsystems ({\it middle}),
 and for an interpolation between them ({\it bottom}). 
For each cases, $p_1^A=p_1^B=p$.}
\end{table}

\begin{figure}[htbp]
\begin{center}
\includegraphics[scale=0.4]{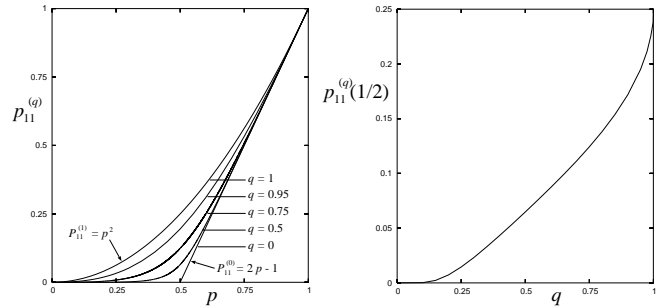}
\end{center}
\caption{$N=2$ : $A=B$ case with $W_A=W_B=2$. {\it Left:} Functions $f_q(p) \equiv p_{11}^{(q)}(p)$ for typical values of $q$. 
{\it Right:} $q$-dependence of $f_q(1/2)$ ($f_1(1/2)=1/4$).}
\end{figure}

\begin{figure}[htbp]
\begin{center}
\includegraphics[scale=0.38]{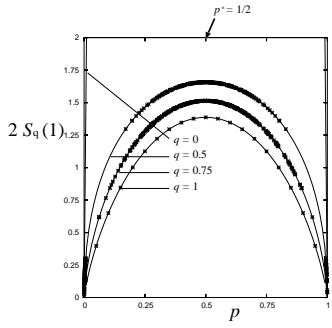}
\end{center}
\caption{$N=2$ . 
$p-$dependences of $2S_q(1)$ (solid curve), and of $S_q(2)$ (dots) for typical values of $q$. For $q=0$ there is a cutoff probability  $p^*=1/2$  
indicated by an arrow. Since numerical calculation of $S_q(2)$ becomes unstable when $p$ is too small, we only show values of $p$ whose corresponding entropies are reliably calculated. For $q=0$, the rest of the curve (i.e., for $0<p<1/2$) can be obtained by using the same solution but permutating $p$ and $1-p$.
}
\end{figure}

\begin{figure}[htbp]
\begin{center}
\hspace{3mm}
\includegraphics[scale=0.4]{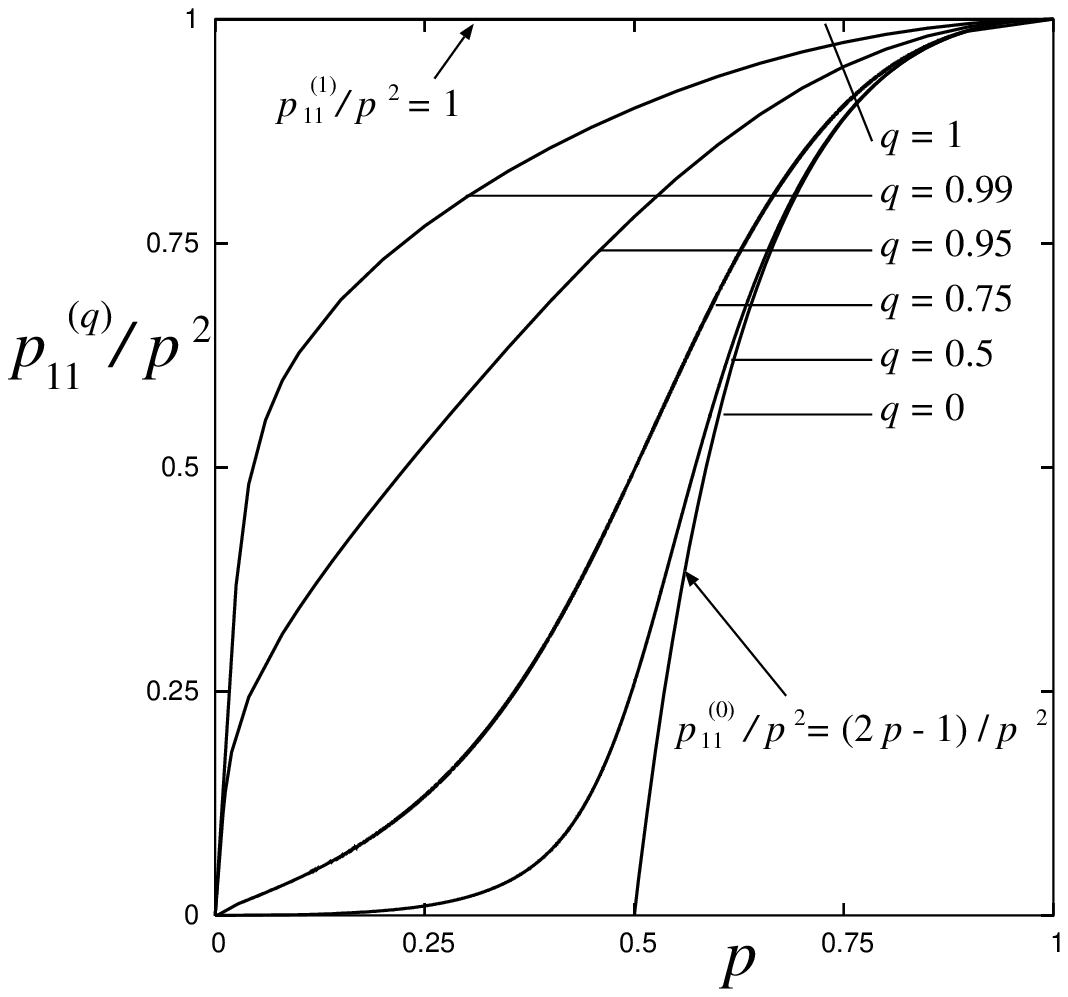}
\end{center}

\begin{center}
\includegraphics[scale=0.4]{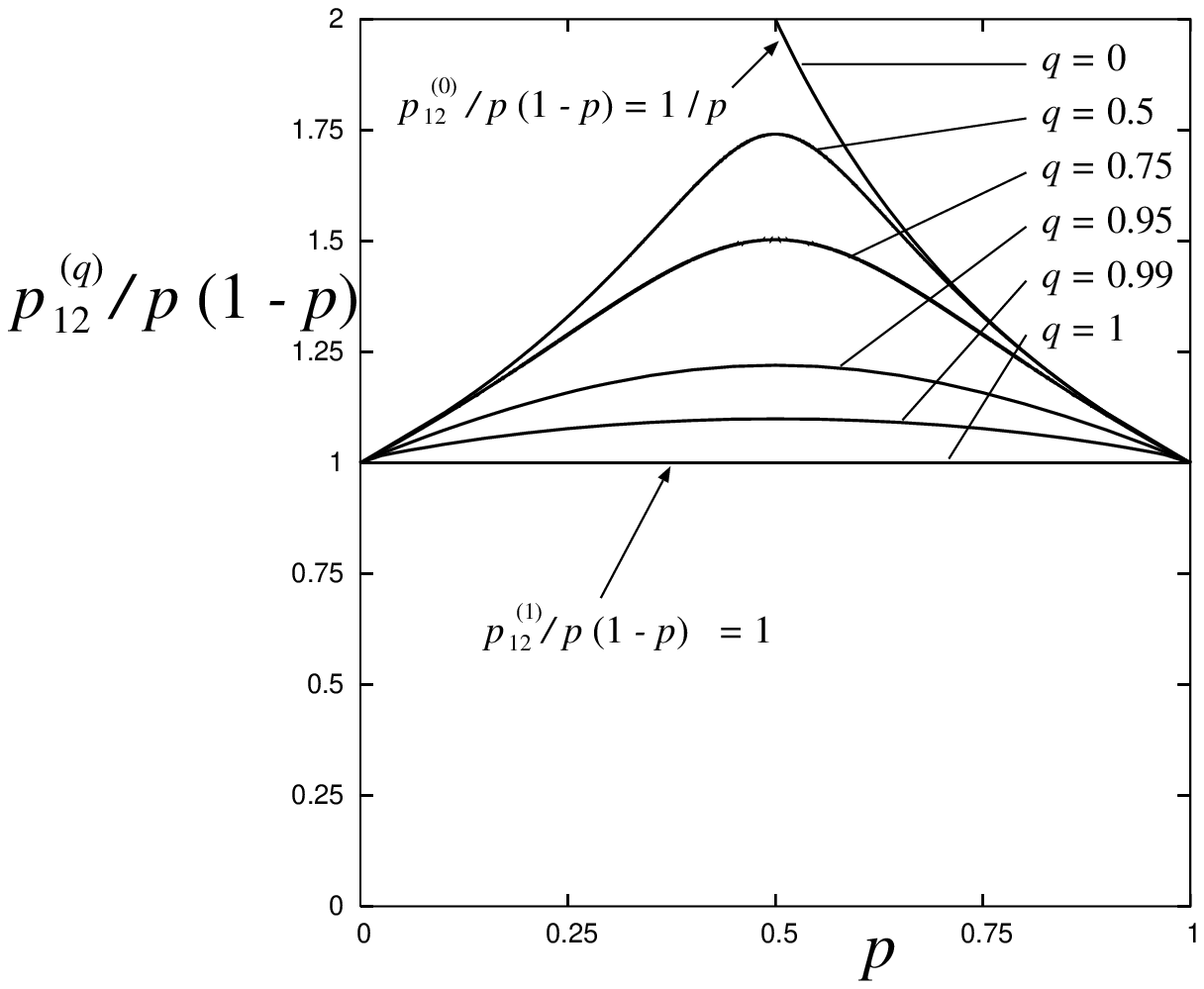}
\end{center}

\begin{center}
\includegraphics[scale=0.4]{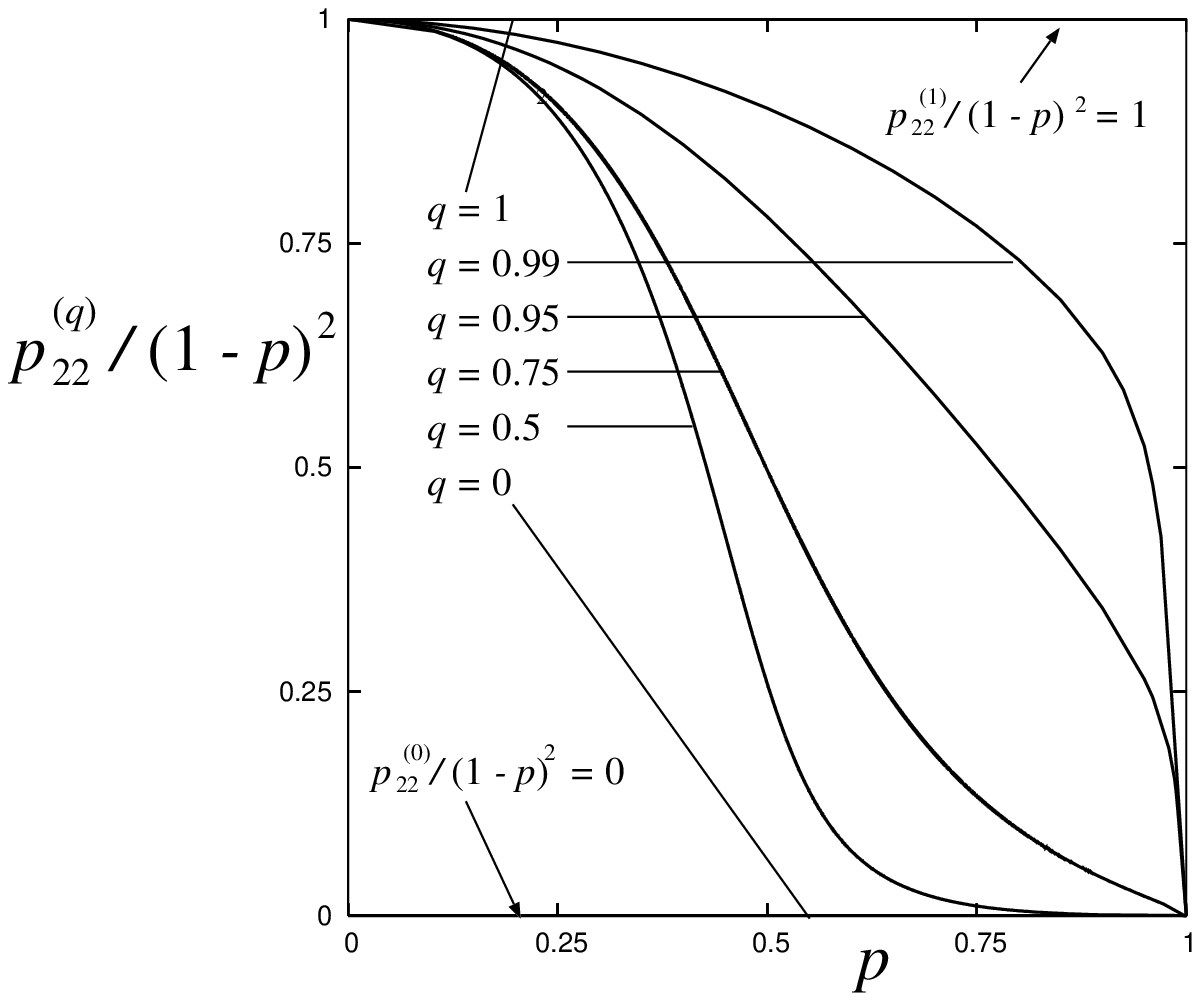}
\end{center}
\caption{$N=2$ . 
$p-$dependence of the normalized ratio $p_{11}^{(q)}/p^2=f_q(p)/p^2$
 ({\it top}), $p_{12}^{(q)}/[p(1-p)]=p_{21}^{(q)}/[p(1-p)]=[p-f_q(p)]/[p(1-p)]  $
 ({\it middle}), 
and $p_{22}^{(q)}/(1-p)^2=[1-2p+f_q(p)]/(1-p)^2  $ ({\it bottom}).}
\label{fig:2norm}
\end{figure} 

It is instructive to see the joint probabilities of this simple case
{\it normalized by those of the independent case}. The results are
shown in Figs. 4. 
\begin{eqnarray}
\mbox{({\it top})}&:&\frac{p_{11}^{(q)}}{p^2}=\frac{f_q(p)}{p^2}, \\
\mbox{({\it middle})}&:&\frac{p_{12}^{(q)}}{p(1-p)}=\frac{p_{21}^{(q)}}{p(1-p)}
=\frac{p-f_q(p)}{p(1-p)}, \\
\mbox{({\it bottom})}&:&\frac{p_{22}^{(q)}}{(1-p)^2}=\frac{1-2p+f_q(p)}{(1-p)^2}. 
\end{eqnarray}
We verify that $q$ decreasing from 1 to zero inhibits the occupation 
of the states $(11)$ and $(22)$, and consistently enhances the occupation of the states $(12)$ and $(21)$. 

\subsection{$N=3$ specially correlated binary systems}

The most general joint probabilities for three binary subsystems
(noted $A$, $B$ and $C$, with $W_A=W_B=W_C=2$) are indicated in Table
V where the numbers without parentheses correspond to
system $C$ in state 1, and the numbers within parentheses correspond 
to system $C$ in state 2.

\begin{table}[htbp]
\begin{center}
\begin{tabular}{c||c|c||}
 $_A\setminus^B$    &  1                                    & 2                                                    \\
[1mm] \hline\hline
1                              &  $\;\;p_{111}^{A+B+C}\;\;$                           & $\;\;p_{121}^{A+B+C}\;\;$    \\   
                                &$(p_{112}^{A+B+C})$                                  &$(p_{122}^{A+B+C})$            \\                                               
[3mm] \hline
2                              &  $p_{211}^{A+B+C}$                                    & $p_{221}^{A+B+C}$             \\
                                &  $(p_{212}^{A+B+C})$                                  & $(p_{222}^{A+B+C})$          \\                                         
[3mm] \hline \hline
\end{tabular}
\end{center}
 \label{tab:jp3bs}
\caption{Joint probabilities for three binary subsystems (the quantities between parentheses correspond to $C$ being in state 2, whereas the others correspond to $C$ being in state 1).}
\end{table}

The corresponding $AB-${\it marginal probabilities} 
are indicated in Table VI which precisely reproduces the situation 
we had for the two-subsystem ($A+B$) problem. 
This is to say $p_{11}^{A+B}=p_{111}^{A+B+C} + p_{112}^{A+B+C}$, $p_{12}^{A+B}=p_{121}^{A+B+C} + p_{122}^{A+B+C}$, 
and so on.

\begin{table}[htbp]
\begin{center}
\begin{tabular}{c||c|c||}
 $_A\setminus^B$    &  1                                    & 2                                                    \\
[1mm] \hline\hline
1                              &  $\;\;p_{11}^{A+B}\;\;$                           & $\;\;p_{12}^{A+B}\;\;$    \\                                                
[3mm] \hline
2                              &  $p_{21}^{A+B}$                                    & $p_{22}^{A+B}$             \\                                       
[3mm] \hline \hline
\end{tabular}
\end{center}
 \label{tab:mp3bs}
\caption{$AB$-marginal probabilities for three binary subsystems}
\end{table}

\begin{table}[htbp]
\begin{center}
\begin{tabular}{c||c|c||}
 $_A\setminus^B$    &  1                                                                   & 2                                                    \\
[1mm] \hline\hline
1                              &  $\;\;p_1^Ap_1^Bp_1^C\;\;$                            & $\;\;p_1^Ap_2^Bp_1^C\;\;$    \\   
                                & $(p_1^Ap_1^Bp_2^C)$                                  &$(p_1^Ap_2^Bp_2^C)$            \\                                               
[3mm] \hline
2                              &  $p_2^Ap_1^Bp_1^C$                                    & $p_2^Ap_2^Bp_1^C$             \\
                                &  $(p_2^Ap_1^Bp_2^C)$                                  & $(p_2^Ap_2^Bp_2^C)$          \\                                         
[3mm] \hline \hline
\end{tabular}
\end{center}

\begin{center}
\begin{tabular}{c||c|c||}
 $_A\setminus^B$    &  1                                    & 2                                                    \\
[1mm] \hline\hline
1                              &  $\;\;p_1^A+p_1^B+p_1^C-2\;\;$                           & $\;\;1-p_1^B\;\;$    \\   
                                & $(1-p_1^C)$                                                        &$(0)$            \\                                               
[3mm] \hline
2                              &  $1-p_1^A$                                                         & $0$             \\
                                &  $(0)$                                  & $(0)$          \\                                         
[3mm] \hline \hline
\end{tabular}
\end{center}

\begin{center}
\begin{tabular}{c||c|c||}
 $_A\setminus^B$    &  1                                                                                        & 2                                             \\
[1mm] \hline\hline
1                              &  $f_q(p_1^A,p_1^C)+f_q(p_1^B,p_1^C)$                            & $-f_q(p_1^A,p_1^B)$                    \\   
                                &  $ -p_1^C(p_1^A+p_1^B)$                                                  & $   +p_1^A(p_1^B+p_1^C)$             \\
                                &  $  +p_1^Cf_q(p_1^A,p_1^B)  $                                         & $ -p_1^Af_q(p_1^B,p_1^C)$                                \\
                                &  $[f_q(p_1^A,p_1^B)+p_1^C(p_1^A+p_1^B)$                      &$[p_1^A(1-p_1^B-p_1^C$            \\      
                                &  $-f_q(p_1^A,p_1^C)-f_q(p_1^B,p_1^C)$                             & $+f_q(p_1^B,p_1^C))]$            \\
                                &  $-p_1^Cf_q(p_1^A,p_1^B)]$                                                &                                                                    \\
[3mm] \hline
2                              &  $-f_q(p_1^A,p_1^B)+p_1^B(p_1^A+p_1^C)$                       & $p_1^C(1-p_1^A-p_1^B$               \\
                                &  $-p_1^Bf_q(p_1^A,p_1^C)$                                                  & $+f_q(p_1^A,p_1^B))$                 \\
                                &  $[p_1^B(1-p_1^A-p_1^C$                                                       & $[(1-p_1^C)(1-p_1^A-p_1^B$             \\     
                                &  $+f_q(p_1^A,p_1^C))]$                                                       & $+f_q(p_1^A,p_1^B))]$            \\                        
[3mm] \hline \hline
\end{tabular}
\end{center}
 \label{tab:jpsc3bs}
\caption{Joint probabilities 
for independent three binary subsystems ({\it top}), 
for specially correlated three binary
 subsystems ({\it middle}), and for an interpolation between them ({\it botom}).}
\end{table}

\begin{table}[htbp]
\begin{center}
\begin{tabular}{c||c|c||}
 $_A\setminus^B$    &  1                                                                & 2                                                    \\
[1mm] \hline\hline
1                              &  $p^3$                                                         & $p^2(1-p)$    \\   
                                & $[p^2(1-p)]$                                                &$[p(1-p)^2]$            \\                                               
[3mm] \hline
2                              &  $p^2(1-p)$                                                 & $p(1-p)^2$             \\
                                &  $[p(1-p)^2]$                                               & $[(1-p)^3]$          \\                                         
[3mm] \hline \hline
\end{tabular}
\end{center}

\begin{center}
\begin{tabular}{c||c|c||}
 $_A\setminus^B$    &  1                                                                & 2                                                    \\
[1mm] \hline\hline
1                              &  $3p-2$                                                        & $1-p$    \\   
                                & $[1-p]$                                                        &$[0]$            \\                                               
[3mm] \hline
2                              &  $1-p$                                                         & $0$             \\
                                &  $[0]$                                      & $[0]$          \\                                         
[3mm] \hline \hline
\end{tabular}
\end{center}

\begin{center}
\begin{tabular}{c||c|c||}
 $_A\setminus^B$    &  1                                                                & 2                                                    \\
[1mm] \hline\hline
1                              &  $2(f_q(p)-p^2)+pf_q(p)$                            & $2p^2-f_q(p)-pf_q(p)$    \\   
                                & $[2p^2-f_q(p)-pf_q(p)]$                               &$[p(1-2p+f_q(p))]$            \\                                               
[3mm] \hline
2                              &  $2p^2-f_q(p)-pf_q(p)$                                & $p(1-2p+f_q(p))$             \\
                                &  $[p(1-2p+f_q(p))]$                                      & $[(1-p)(1-2p+f_q(p))]$          \\                                         
[3mm] \hline \hline
\end{tabular}
\end{center}
 \label{tab:jpsc3bss}
\caption{Joint probabilities for independent three binary subsystems
 ({\it top}), for specially correlated three binary subsystems({\it middle}), and 
for an interpolation between them ({\it bottom}). 
For each cases, $p_1^A=p_1^B=p_1^C=p$}
\end{table}

\begin{figure}[htbp]
\begin{center}
\includegraphics[scale=0.38]{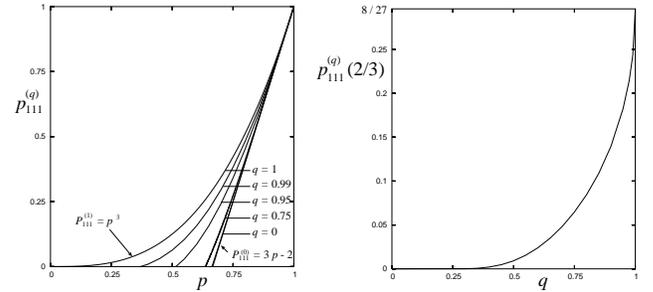}
\end{center}
\caption{$N=3$ . {\it Left:} $p-$dependence of $p_{111}^{(q)}$ for typical values of $q$. {\it Right:} $q-$dependence of $p_{111}^{(q)}(2/3)$ $(p_{111}^{(1)}(2/3)=8/27)$.
}
\end{figure}

\begin{figure}[htbp]
\begin{center}
\includegraphics[scale=0.38]{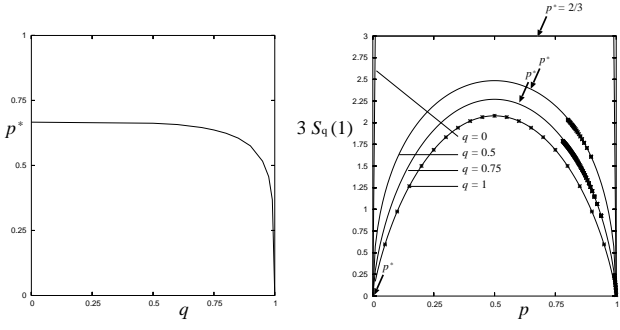}
\end{center}
\caption{$N=3$ . {\it Left:} $q$-dependence of the cuttoff probability $p^*$. {\it Right:} 
$p-$dependences of $3S_q(1)$ (solid curve), and of $S_q(3)$ (dots) for typical values of $q$. 
The corresponding cutoff probabilities $p^*$ are 
indicated by arrows. Since numerical calculation of $S_q(3)$ 
becomes unstable when $p$ is close to  $p^*$, we only show values of $p$ not too close to $p^*$. For all values of $q$ such that $p^*(q)<1/2$, we can easily obtain the low-$p$ branch of $S_q(3)$ just by using the permutation $p$ with $1-p$. However, the situation is considerably more complex for those values of $q$ such that $1/2<p^*(q) \le 2/3$; in this case a new branch of solutions is needed to cover the region $1-p^*<p<p^*$. We have not addressed this situation.}
\end{figure}

\begin{figure}[htbp]
\begin{center}
\hspace{6mm}
\includegraphics[scale=0.4]{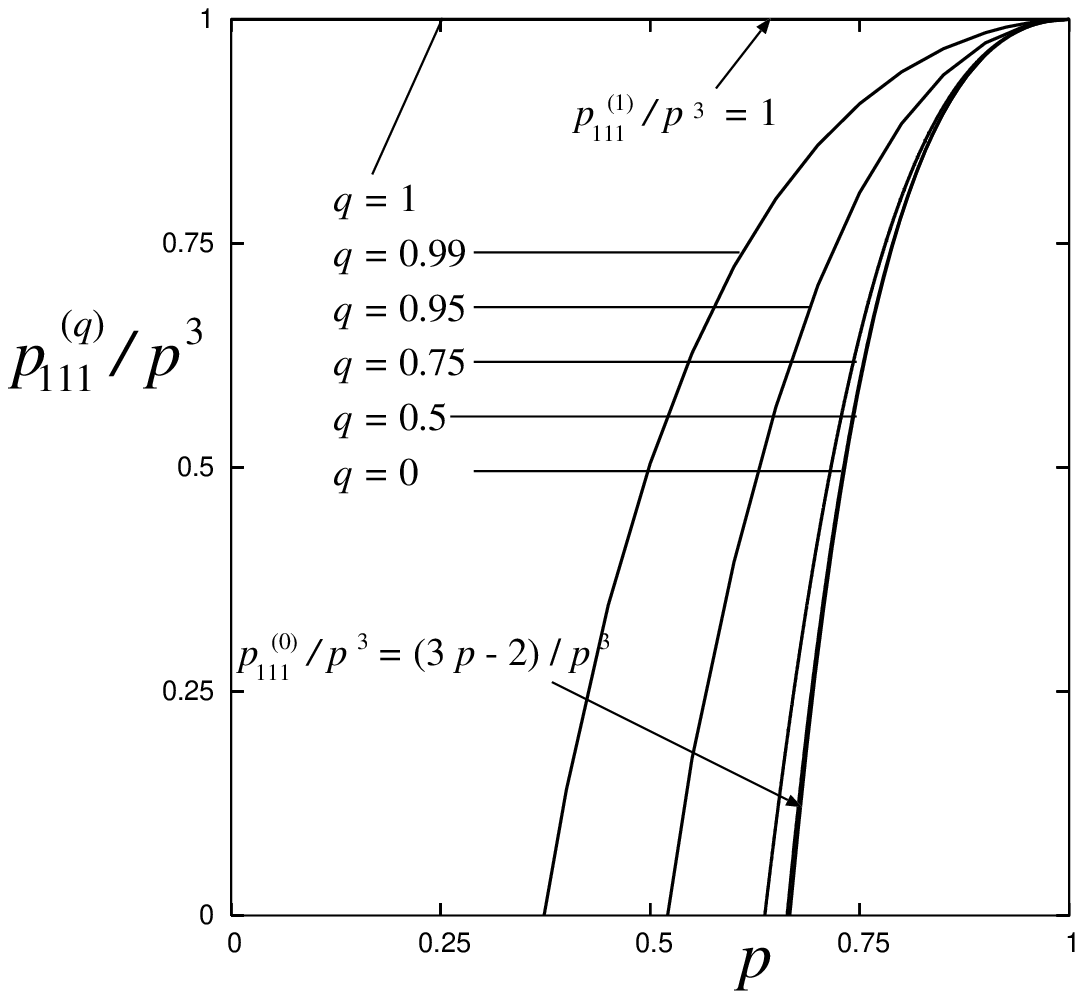}
\end{center}

\begin{center}
\includegraphics[scale=0.4]{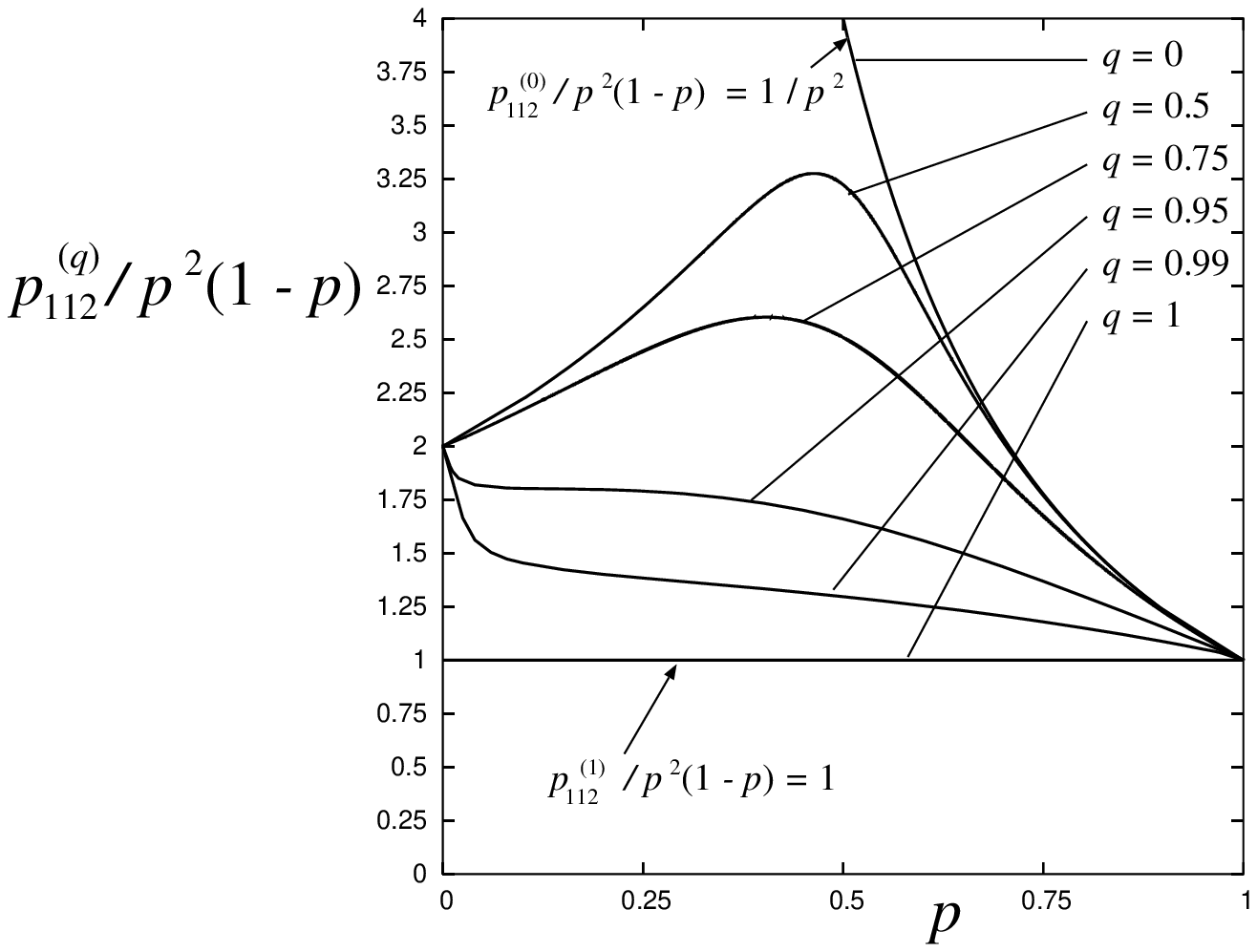}
\end{center}

\begin{center}
\includegraphics[scale=0.4]{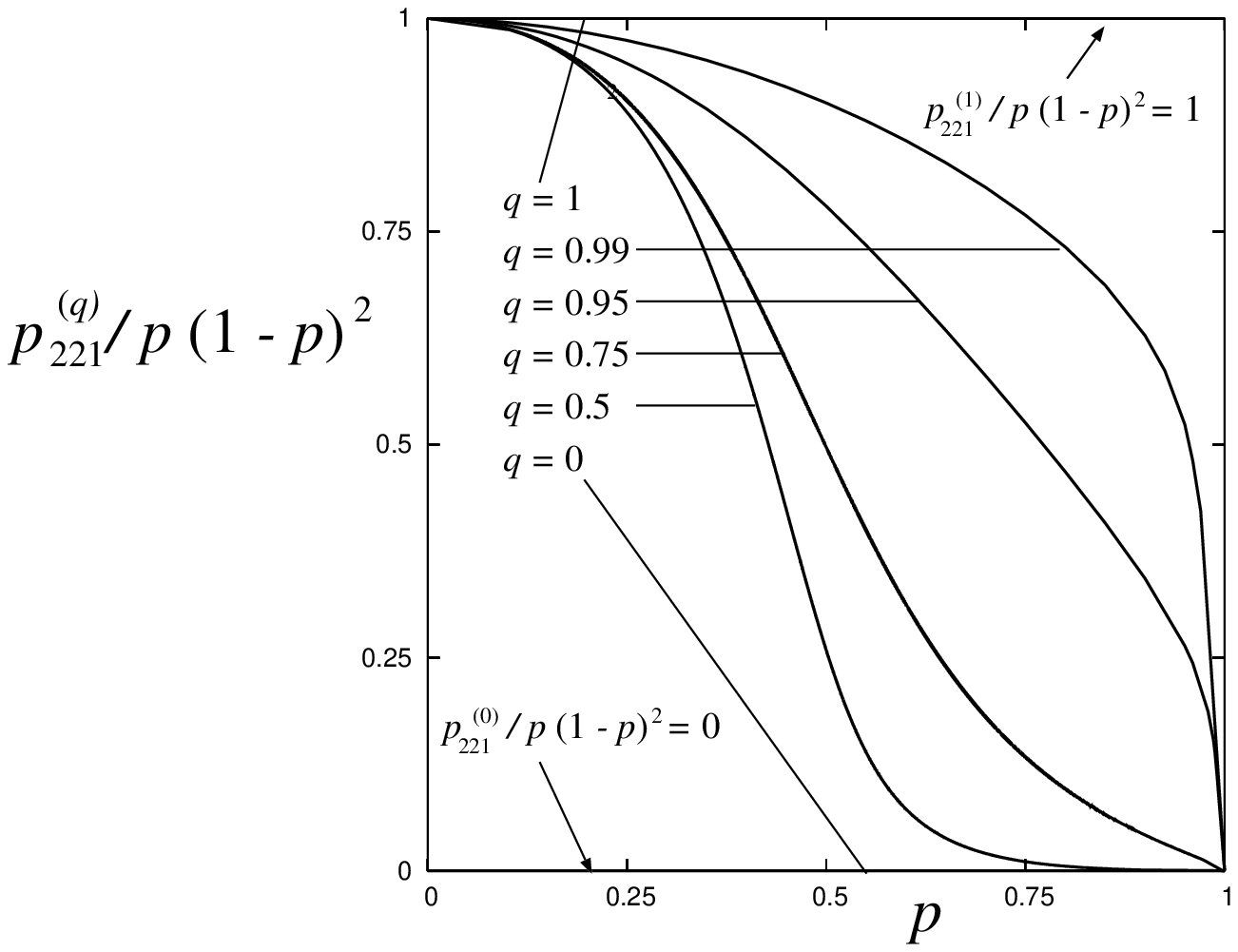}
\end{center}
\caption{$N=3$ . 
$p-$dependence of the normalized ratio $p_{111}/p^3=
 [2(f_q(p)-p^2)+pf_q(p)]/p^3$  ({\it top}), 
$p_{121}^{(q)}/[p^2(1-p)]=    p_{211}^{(q)}/[p^2(1-p)] =p_{112}^{(q)}/[p^2(1-p)]  =
 [2p^2-f_q(p)-pf_q(p)]/[p^2(1-p)] $ ({\it middle}), and 
$p_{221}^{(q)}/[p(1-p)^2]= p_{122}^{(q)}/[p(1-p)^2]=p_{212}^{(q)}/[p(1-p)^2]=pn_{222}^{(q)}/(1-p)^3=   (1-2p+f_q(p)/(1-p)^2$ ({\it bottom}).
}
\end{figure}

The mutually independent case, specially correlated case   (with $p_1^A + p_1^B + p_1^C > 2$), 
and  an interpolation between them are indicated in Table VII. We verify Eq. (17) for $q=0$ and $q=1$.
The $AB-${\it marginal probabilities} recover Table II. 
For the simple particular case $A=B=C$, the Tables VII  
become respectively the Tables VIII where we have used $f_q(p,p)=f_q(p)$. 

In Fig. 6 we present $p_{111}^{(q)}(p)=2[f_q(p)-p^2]+pf_q(p)$ for typical
values of $q$ and the $q-$dependence of $p_{111}^{(q)}(2/3)$. 
And in Fig. 6 we exhibit the $q-$dependence of the cutoff probability 
$p^*$ below which the $p_{111}$ probability vanishes. 
Finally, in Figs. 7 we present the relevant normalized ratios 

\begin{eqnarray}
\mbox{({\it top})}&:&\frac{p_{111}^{(q)}}{p^3}= 
\frac{2(f_q(p)-p^2)+pf_q(p)}{p^3}, \\
\mbox{({\it middle})}&:&
\frac{p_{121}^{(q)}}{p^2(1-p)}=
\frac{p_{211}^{(q)}}{p^2(1-p)}=
\frac{p_{112}^{(q)}}{p^2(1-p)}    \nonumber\\
&=&\frac{2p^2-f_q(p)-pf_q(p)}{p^2(1-p)}, \\
\mbox{({\it bottom})}&:&
\frac{p_{221}^{(q)}}{p(1-p)^2}= 
\frac{p_{122}^{(q)}}{p(1-p)^2}=
\frac{p_{212}^{(q)}}{p(1-p)^2}\nonumber\\
&=&\frac{p_{222}^{(q)}}{(1-p)^3} 
=\frac{1-2p+f_q(p)}{(1-p)^2}.
\end{eqnarray}

\section{Final comments and conclusions}

The so called ``nonextensive" entropy $S_q$ ($q \ne 1$) is in fact {\it nonextensive} ({\it nonadditive} strictly speaking) if we are composing subsystems $A$ and $B$ assumed  (sometimes {\it explicitly}, but most of the times {\it tacitly}) to be independent. Indeed, in such case we have $S_q(A+B)=S_q(A)+S_q(B)+(1-q)S_q(A)S_q(B)$, which, unless $q=1$, generically differs from $S_q(A)+S_q(B)$. It is from this property that currently used expressions such as {\it nonextensive entropy} and {\it nonextensive statistical mechanics} stand. However, special types of collective correlations do exist for which extensivity is recovered for the appropriate value of $q$. This is to say, correlations such that $S_q(A+B)=S_q(A)+S_q(B)$. This situation has been illustrated for $N=2$ and $N=3$ equal binary subsystems in Figs. 3 and 6 respectively. 

The $q=0$ case has also been illustrated (see \cite{tsallisSF}) for an arbitrary number $N$ of arbitrary subsystems (with number of states $W_1,W_2,...,W_N$ respectively). The a priori total number of states is $W=\prod_{r=1}^N W_r$, but most of them have zero probability. In other words, the number of effective states (i.e., those whose probability is generically nonzero) is only $W^{\mbox{\it eff}}=(\sum_{r=1}^N W_r)-N+1$. Consequently, a necessary condition for this very special type of correlation to occur is the system to have  $W-W^{\mbox{\it eff}}=(\prod_{r=1}^N W_r)  - (\sum_{r=1}^N W_r)+N-1$  zeros in its table of joint probabilities. If the subsystems are all equal, we have $W=W_1^N$, whereas $W^{\mbox{\it eff}}=N(W_1-1)+1$. At the thermodynamic limit, it clearly is $W^{\mbox{\it eff}}<<W$, i.e., $\lim_{N\to\infty}[W^{\mbox{\it eff}}(N)/W(N)]=0$. If our subsystems were such that $W_r \to\infty\;(\forall r)$ yielding a {\it continuum}, they would ultimately lead to a {\it finite} Lebesgue measure. This measure would be the hypervolume $W$ associated with $Nd_1$ dimensions, i.e., essentially $W=\prod_{r=1}^N L_r$ where $L_r$ is the Lebesgue measure associated with the $d_1-$dimensional $r-$th subsystem. In remarkable contrast, $W^{\mbox{\it eff}}$ would correspond to a set of {\it zero} Lebesgue measure, such as, for instance, a (multi)fractal whose Hausdorff dimension $d_H$ would be {\it smaller} than $Nd_1$. 

Within such a scenario, it is natural to conjecture the following situation for $q$ decreasing from say 1 to 0 (see also \cite{tsallisSF}). When the subsystems are strictly independent (no correlations at all) or nearly independent (typically {\it short} range two-body interactions within a many-body Hamiltonian system), we expect an {\it exponential} $N$-dependence $W \sim W^{\mbox{\it eff}} \sim \mu^N \;(\mu\ge 1)$, and the extensive entropy to be the BG one. In contrast, if the subsystems have special collective correlations (typically {\it long} range two-body interactions within a many-body Hamiltonian system), we expect a {\it power-law} behavior $W^{\mbox{\it eff}} \sim N^{\rho}\;(\rho \ge 0) <<W$, and the extensive entropy to be $S_q$ with $q=1-1/\rho$. Consistently, if $\rho=1$, then the extensive entropy is $S_0$. For all $q<1$ we expect, in the continuum case, a zero Lebesgue measure, and a fractal dimension $d_H$ decreasing with decreasing $q$. A pictorial image can help understanding the {\it conjecture}. Travelling in a Brownian way --- say an hypothetical ``blind" cowboy on the back of an hypothetical ``blind" horse --- will lead to a virtually homogeneous visit of a big (relatively plane) territory, associated with a {\it finite} Lebesgue measure. And the result would roughly be the same independently of the initial condition (starting point of the travel). This would be the typical dynamics associated with {\it strong} chaos (i.e., {\it positive} Lyapunov exponents), thus leading to the $q=1$ entropy.  Travelling in the way of a pilot of an airline company is quite different. First of all, he(she) will only visit the set of airports through which this company operates. Given the small size of the airports (compared to the size of a wide territory), this set constitutes a set of virtually {\it zero} Lebesgue measure. Although statistically similar in geometrical terms, the result does depend on the initial conditions (the most important hub of the network of Japan airlines is Tokyo, whereas the most important of Varig is Sao Paulo, and the most important one of Continental airlines is Houston). Although no rigorous proof whatsoever is yet available,
the typical dynamics to be associated is {\it expected} to be that of {\it weak} chaos (i.e., basically {\it zero} Lyapunov exponents). We expect the adequate entropy to be $q \ne 1$. If we wish to recover homogeneity in the visits, we need to average over virtually all the possible initial conditions. Such an average is not needed for the $q=1$ case. 

Occupancy of a phase space without strong restrictions makes equal probabilities of the joint system compatible with equal probabilities of each subsystem. This compatibility disappears if visiting of some regions of the joint space is strongly enhanced whereas visiting of others is strongly inhibited (see Figs. 4 and 7). This is clearly seen for the generic $q=0$ case for any value of $N$. It might be necessary to go to the asymptotic $N\to\infty$ limit in order to see it for say $0<q<1$.  In any case, the dynamics conjectured for the $q \ne 1$ cases seem to be compatible with a recent connection \cite{carati} in terms of recurrent visits in phase space (the $q=1$ limit corresponding to a Poisson distribution of times between consecutive visits, during the time evolution of the whole system, of a given cell of phase space).

Let us elaborate some more on the important connection of $q$ with geometry. We consider, for simplicity, the case of $N$ equal subsytems $A_1,A_2,...,A_N$, each of them having $W_1$ possible microstates. The total space has then $W=W_1^N$ possible microstates that can be represented on a discrete $N$-dimensional hypercube of linear size $W_1$. We shall focus however on the effective number $W^{\mbox{\it eff}} \;(W^{\mbox{\it eff}}\le W)$ of microstates whose probability generically is {\it not} zero. A most trivial occupation is when only the ``origin" corner of the hypercube is occupied, i.e., $p_{11...1}^{A_1,A_2,...,A_N}=1$. We then have $W^{\mbox{\it eff}}=1$. A simple nontrivial case is when , in addition to that ``corner", the $N$ ``edges" of the hypercube starting from the ``corner" are occupied as well (with generic probabilities). We then have $W^{\mbox{\it eff}}=1+N(W_1-1)$, hence $\rho=1$ and $q=0$, as already addressed. A next case in this series is occupancy of all the $N(N-1)/2$ ``faces" starting at that ``corner". We then have $W^{\mbox{\it eff}}=1+N(W_1-1)+[N(N-1)/2](W_1-1)^2$, hence $\rho=2$ and $q=1/2$. 
The next case in this series is occupancy of all the $N(N-1)(N-2)/6$ ``cubes" starting at that ``corner". We then have $W^{\mbox{\it eff}}=1+N(W_1-1)+[N(N-1)/2](W_1-1)^2+[N(N-1)(N-2)/6](W_1-1)^3$, hence $\rho=3$ and $q=2/3$. In general, if all the $r$-dimensional hypercubes starting at that ``corner" are occupied, we have  $W^{\mbox{\it eff}}=1+N(W_1-1)+[N(N-1)/2](W_1-1)^2+[N(N-1)(N-2)/6](W_1-1)^3+...+[N!/(N-r)!r!](W_1-1)^r$, hence $\rho=r$ and $q=(r-1)/r=1-1/r$. The last element of this series corresponds to fully occupy the unique $N$-dimensional hypercube. 
We then have  $W^{\mbox{\it eff}}=1+N(W_1-1)+[N(N-1)/2](W_1-1)^2+...+[N!/(N-r)!r!](W_1-1)^r+...+N(W_1-1)^{N-1}+(W_1-1)^N=[1+(W_1-1)]^N=W_1^N$, hence $q=1$. A different geometry which nevertheless belongs to the $q=1/2$ universality class is the following: if, in addition to the ``corner", all the diagonals of the $N(N-1)/2$ ``faces" starting at that ``corner" also are occupied, we have $W^{\mbox{\it eff}}=1+[N(N-1)/2](W_1-1)$, hence $\rho=2$ and $q=1/2$. Another different geometry could be the following one: assume that, in addition to the ``corner", the ``edges that are (strictly or substantially) occupied are not all $N$ edges, but only those following the Cantor set sequence $101000101000000000101000101...$, whose fractal dimension is $d_H^{(N)}=\ln 2/\ln3$. We then have $W^{\mbox{\it eff}}=1+N^{d_H^{(N)}}(W_1-1)$ (we are assuming that $N$ is a power of $3$), hence $\rho=d_H^{(N)}<1$ and $q=1-1/d_H^{(N)}<0$. A similar situation can occur for the $W_1$ states. Suppose that, in addition to the ``corner", all the $N$ ``edges" are occupied but {\it not} fully occupied. Suppose that the $(W_1-1)$ states are fractally occupied (again the Cantor, or any other sequence) with fractal dimension $d_H^{(W)}$. We then have $W^{\mbox{\it eff}}=1+N(W_1-1)^{d_H^{(W)}}$, hence $\rho=1$ and $q=0$. A quite general situation could be, in the $(N,W_1) \to (\infty,\infty)$ limit,  $W^{\mbox{\it eff}} \sim N^{d_H^{(N)}} W_1^{d_H^{(W)}}$. In all these illustrations, the probabilities associated with the $W^{\mbox{\it eff}}$ occupied microstates have no particular reason for being {\it equally} probable. They could very well constitute a network (e.g., a scale-free network) whose occupancy probabilities would characterize main hubs, and secondary hubs, and so on (quite like the previously mentioned set of airports used by a given airlines company). In fact, it would not be really surprising if classical long-range-interacting many-body Hamiltonian systems would visit cells in phase space according to probabilities of this type. 

This is the typical scenario we expect for the family of entropies $S_q$. It is easy to imagine that there can easily be even more subtle situations for which the apropriate (extensive) entropy would be {\it not} included in the $S_q$ family for {\it any} value of $q$. Different entropic forms would perhaps be then needed. But even within the $S_q$ family, various aspects remain to be solved, that have been only preliminarly addressed here. Let us mention two of them. 

First, the solutions that we have exhibited here are probably not unique (see the captions of Figs. 3 and 6). Other branches of solutions could well exist. We have been unable, at the present stage, to find them all. The reader surely realizes the nontrivial mathematical difficulty of {\it simultaneously} satisfying the impositions of theory of probabilities (sum of all the joint probabilities equal to unity, partial sums of the joint probabilities equal to the marginal probabilities) and those of extensivity of $S_q$. The general solution seems to be \cite{tsallisSF} intimately related with the recently introduced $q$-product \cite{borges} $x \times_q y \equiv (x^{1-q}+y^{1-q}-1)^{1/(1-q)}$, which has, among others, the following properties: (i) $x \times_1 y=xy $, (ii) $\ln_q (x \times_q y)=\ln_q x+ \ln_q y$ (whereas $\ln_q (x y)=\ln_q x+ \ln_q y+(1-q)\ln_q x \ln_q y)$; (iii) $1/(x \times_q y)=(1/x) \times_{2-q}(1/y)$; (iv) $x \times_q(y \times_q z)=(x \times_q y) \times_q z=x \times_q y \times_q z= (x^{1-q}+y^{1-q}+z^{1-q}-2)^{1/(1-q)}$; (v) $x \times_q 1=x$. This interesting structure probably is one of the ingredients, but there are surely others to be considered concomitantly, specifically those related to the constraints imposed by theory of probabilities.

Second, to illustrate an important point let us rewrite Eq. (22) as follows:
\begin{equation} 
2p^\kappa+2(1-p)^\kappa-(f_\kappa)^\kappa-2(p-f_\kappa)^\kappa-(1-2p+f_\kappa)^\kappa=1\,.
\end{equation}   
This relation means that it has been possible to find a function $f_\kappa(p)$ which satisfies the impositions of theory of probabilities. Now, if we wish this solution to correspond to the extensivity of $S_q$, we just impose $\kappa=q$. If we wish instead to impose the extensivity of $S_{2-q}$ we identify $\kappa=2-q$. In this case, we have solutions corresponding to $1 \le q \le 2$. We can even impose, if we wish, the extensivity of $S_{\kappa(q)}$, where $\kappa(q)$ is virtually any (increasing or decreasing) monotonic function of $q$ satisfying $\kappa(1)=1$. This freedom {\it might} play a relevant role when a thermodynamical (or thermodynamical-like) connection  is seeked. Indeed, most of the systems which seem to obey nonextensive statistical mechanics exhibit a (quasi)stationary state whose entropic index is $q \ge 1$. This point needs further analysis in order to unambiguosly establish the identification between $\kappa$ and $q$ which is thermodynamically adequate. It is interesting at this stage to recall a recent discussion by Robledo \cite{robledomori} on a nonthermodynamical system, which has nevertheless some analogy with the present situation. Basically, the so called Mori's {\it $q$-transitions} for say the usual logistic map occur at both .2445... and 2 - 0.2445... \cite{mori}.

A few words on terminology to conclude. We have seen that (under specially correlated composition of subsystems) $S_q$ can be strictly additive (i.e., $S_q(A+B)=S_q(A)+S_q(B)$) for a variety of values of the entropic index $q$. It has nevertheless become current denomination to refer to  the $q=1$ universality class as {\it extensive} or {\it normal} systems, and to the $q \ne 1$ universality classes as {\it nonextensive} or {\it anomalous} systems. This use has originated from the fact that, in the early times of the theory, the focus was explicitly or tacitly put onto independent subsystems. {\it For this simple composition law, and only then, $S_{BG}$ is strictly additive (i.e., $S_{BG}(A+B)=S_{BG}(A)+S_{BG}(B)$), whereas $S_q$ ($q \ne 1$) is not (i.e., $S_q(A+B)=S_q(A)+S_q(B)+(1-q)S_q(A)S_q(B)$)}. Although slightly misleading from the entropy standpoint, the current notational distinction {\it extensive} versus {\it nonextensive} is instead perfectly natural  from the energy standpoint of Hamiltonian systems. Indeed, suppose we have a $d$-dimensional classical system with attractive  two-body interactions whose potential energy decays with (dimensionless) distance $r$ as $-1/r^\alpha \;(\alpha \ge 0)$. Let us further assume for simplicity that a strong repulsion exist at the $r \to 0$ limit (therefore no nonintegrable singularities exist at short distances). The Lennard-Jones gas would be $(\alpha,d)=(6,3)$; Newtonian gravitation  would be $(\alpha,d)=(1,3)$ (if we take into account the fact that at very short distances, important repulsive quantum effects are expected which would avoid the mathematical problems tied to the $-1/r$ singularity). The total potential energy at the ground state is expected to be $U(N) \propto -N \int_{1}^{N^{1/d}} dr\, r ^{d-1} r^{-\alpha} \propto N[N^{1-\alpha/d}-1]/(1-\alpha/d)$, where we have assumed for simplicity that the $N$ elements of the system are roughly homogeneously distributed in space, and where the dimensionless distance $r$ has been chosen to be unity at the short distance effective cutoff. We immediately see that the energy is extensive if $\alpha/d>1$, whereas it is nonextensive if $0 \le \alpha/d \le1$. It is long known (see, for instance, \cite{fisher}) that, for the $\alpha/d>1$ systems, BG statistical mechanics is perfectly adequate. More over, for them the $t\to\infty$ and the $N\to\infty$ limits commute, thus always leading to thermal equilibrium. On the other hand, plethoric evidence now exists that, in remarkable variance, for the $0 \le \alpha/d \le 1$ systems, the $t\to\infty$ and the $N\to\infty$ limits do not commute, the physically interesting states for large systems being the (quasi)stationary or metastable ones corresponding to taking first the $N\to\infty$ limit and only afterwards the $t\to\infty$ limit. For such anomalous states, the inadequacy of BG statistical mechanics is notorious {\it when we use no other dynamics than the natural one} (Newton's law if the system is classical) \cite{aging}. For instance, the distribution of velocities is seemingly {\it not} Maxwellian, and there is {\it aging}, a phenomenon absolutely incompatible with the translational invariance expected for thermal equilibrium. A transparent proof that nonextensive statistical mechanics (with a $q \ne 1$ entropy $S_q$) is in place is still lacking, but this could be the case. Indeed, vanishing Lyapunov exponents have been exhibited, as well as the specific anomalous diffusion associated with the nonlinear Fokker-Planck equation, and a variety of $q-$exponential behaviors \cite{SalinasTsallis}. Work is in progress and further contributions are welcome. 

\section*{Acknowledgments}
We have benefited from useful remarks by M. Gell-Mann and J.D. Farmer.

\end{document}